# Mechanically-reconfigurable van der Waals devices via low-friction gold sliding

Andrew Z. Barabas*[1], Ian Sequeira*[1], Yuhui Yang[1], Aaron H. Barajas-Aguilar[1], Takashi Taniguchi[2], Kenji Watanabe[2], Javier D. Sanchez-Yamagishi[1]

[1]Department of Physics and Astronomy, University of California, Irvine, Irvine, CA, USA
[2]Research Center for Functional Materials, National Institute for Materials Science, 1-1 Namiki, Tsukuba, Japan
*These authors contributed equally to this work

**Abstract**

Interfaces of van der Waals (vdW) materials such as graphite and hexagonal boron nitride (hBN) exhibit low-friction sliding due to their atomically-flat surfaces and weak vdW bonding. We demonstrate that microfabricated gold also slides with low friction on hBN. This enables the arbitrary post-fabrication repositioning of device features both at ambient conditions as well as *in-situ* to a measurement cryostat. We demonstrate mechanically-reconfigurable vdW devices where device geometry and position are continuously-tunable parameters. By fabricating slidable top gates on a graphene-hBN device, we produce a mechanically-tunable quantum point contact where electron confinement and edge-state coupling can be continuously modified. Moreover, we combine *in-situ* sliding with simultaneous electronic measurements to create new types of scanning probe experiments, where gate electrodes and even entire vdW heterostructures devices can be spatially scanned by sliding across a target.

**Introduction**

Nanoscale electronic devices are typically static, with the material structure and device geometry set during the fabrication process. Exploring the full parameter space requires fabricating multiple devices with varying geometries and material structures. Ideally, a device's material structure and geometry would be reconfigurable *in-situ*, allowing for post-fabrication modification while simultaneously measuring its properties. Micro-electro-mechanical systems (MEMS) enable a limited range of mechanical reconfigurability at the cost of complex suspended device structures (*1*). For conventional non-suspended devices, mechanical-modification of the device structure is typically not possible due to high friction forces at all interfaces.

An exception to this is van der Waals (vdW) layered materials, which exhibit low interfacial friction due to weak vdW bonds, atomically-flat layers, and lattice incommensurability (*2–8*). Recently, this property has been exploited to perform twist-angle dependent studies of graphite and graphene-based heterostructures by sliding vdW flakes with an atomic force microscope (AFM) (*9–12*). This approach is powerful, but currently limited by difficulties in fabricating complex vdW heterostructures, as well as the need to perform experiments in ambient conditions.

Here, we show that microfabricated gold exhibits low-friction sliding on hexagonal boron nitride (hBN), a vdW material, at both ambient conditions and at cryogenic temperatures (7.6 K). The low-friction gold-hBN interface enables us to produce a wide-range of slidable structures to form mechanically-reconfigurable vdW devices, including a tunable graphene quantum point contact and sliding-based scanning probe devices. Such devices can be modified ex-situ in an AFM or *in-situ* in a measurement cryostat.

**Results**
**Low friction gold on hexagonal boron nitride**

To create reconfigurable structures, we deposit gold microstructures directly onto hBN flakes using electron beam lithography and electron-beam evaporation (see supplementary information (SI)). By pushing laterally on the gold with an AFM tip, we can easily slide microscale features as large as 35 µm$^2$ across the hBN surface. The low-friction sliding enables arbitrary repositioning of deposited features (Figure 1a-b). We observe that small features can even be moved by scanning an AFM tip in tapping mode. The motions are non-destructive, with no change to either the gold or hBN observable in AFM except for the cleaning of contaminants on the hBN surface, which are swept away by the sliding gold (Figure 1c).

To characterize the friction, we slide gold squares of different sizes on hBN using an AFM tip and determine the interfacial friction from AFM deflection measurements, similar to previous vdW tribological studies (*4–6*, *8*, *13*, *14*). Figure 1d illustrates the friction measurement scheme. First the tip is moved laterally at a fixed z-piezo extension elevated above the hBN surface (left panel). Once the tip makes contact with the edge of the stationary gold square it deflects laterally, resulting in a voltage signal on the AFM photodiode (middle panel). The lateral deflection increases until the static friction of the gold-hBN interface is overcome, after which it drops to a constant value corresponding to the kinetic friction as the gold slides on the hBN (right panel). These regions are highlighted in an example deflection trace in Figure 1e, where the peak (static) deflection voltage and constant (kinetic) voltage are indicated.

We repeat these measurements multiple times each for 0.5 to 3 µm wide gold squares and find that the deflection voltages scale linearly versus area, with slopes of 82 ± 6 mV/µm$^2$ (static) and 24.4 ± 0.6 mV/µm$^2$ (kinetic). Assuming the force is directly proportional to deflection voltage, our data shows a linear scaling of interface friction with area, which is typical for most conventional interfaces (*3*). By contrast, atomically flat and lattice incommensurate interfaces, such as single crystal gold nanoparticles on graphite, can exhibit sublinear scaling versus area (*3*, *4*, *8*, *14*). The sublinear friction scaling versus area expected for single crystal interfaces suggests that increased grain size in our polycrystalline gold will result in lower friction. To test this, we annealed our samples at 350 °C for 30 minutes and observed the deflection voltages decrease by 50%. The annealing process also caused the average gold grain size to increase from ~20 nm to ~80 nm, suggesting a connection; although, the removal of contaminants from the Au-hBN interface by heat annealing likely plays a role as well.

To determine the static and kinetic friction values, we convert from deflection voltage to lateral force using a linear model which requires both the AFM tip's lateral spring constant and the AFM's lateral sensitivity. The spring constant is the ratio of lateral force applied to the tip and lateral displacement, which we determine by simulating our tip in COMSOL Multiphysics. The sensitivity is the ratio of lateral tip displacement and lateral deflection voltage measured on the AFM photodiode. To measure the lateral sensitivity, we use the slope of the static region of our deflection linetraces (see SI for more details).

Converting the deflection voltages yields forces of 6 μN (static) and 2 μN (kinetic) for a 9 µm$^2$ gold square on hBN. Applying the conversion to the linear fits results in a friction force per unit area. For the unannealed Au-hBN interface they are 800 nN/µm$^2$ (static) and 230 nN/µm$^2$ (kinetic), and after annealing they decrease to 400 nN/µm$^2$ and 100 nN/µm$^2$, respectively. These force values have uncertainties of 30%, dominated by uncertainty in the sensitivity, and should be considered upper bounds due to the calibration method and limitations of the linear model (see SI). These interfacial friction values are comparable to prior tribology studies of gold on graphite, which measured 50 to 430 nN/µm$^2$ (kinetic) for ~60 to ~100 nm wide, single crystal, gold nanoparticles (*13*). For additional comparison, the previously reported kinetic friction of unaligned graphite on hBN is smaller, at 15 nN/µm$^2$ (*5*). We have also made friction measurements of gold with a 3 nm Cr sticking layer on hBN, and initial tests show it exhibits roughly an order of magnitude higher friction than annealed gold on hBN without a sticking layer (see SI).

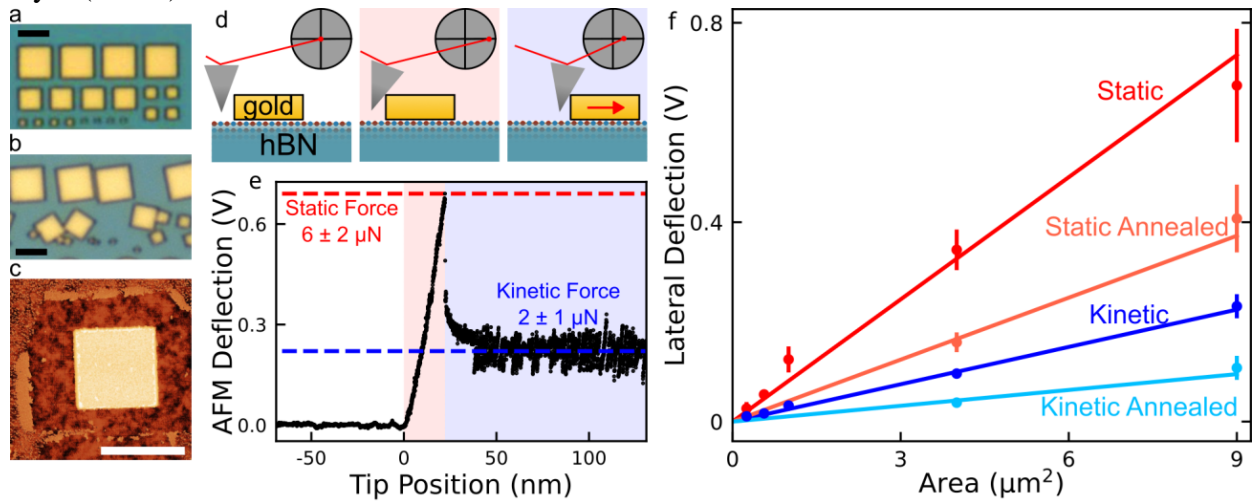

**Figure 1: AFM friction measurements for gold squares sliding on hBN.**
All scale bars are 3 µm. **(a&b)** Optical images of ~170 nm tall gold squares on hBN before and after manipulation with an AFM tip. **(c)** AFM height image of 3 µm wide gold square on atomically flat hBN surface with contaminants swept aside by sliding. **(d)** Schematic illustrating AFM lateral friction measurement: before contact (left), during static friction (middle), during kinetic friction (right). **(e)** AFM friction linetrace for a 3 µm wide square with the tip moving at 1 nm/s. The peak voltage corresponds to the Au-hBN static friction, and the subsequent constant voltage corresponds to the kinetic friction. **(f)** Lateral deflection voltage versus interface area between gold and hBN for 0.5, 0.75, 1, 2 and 3 µm wide squares, before and after annealing at 350 °C for 30 minutes. Each data point is the average of multiple measurements for each size. Error bars show standard deviation. Lines are linear fits through zero, fitting only the 4 and 9 µm$^2$ data points; this excludes smaller deflection data points which have variable AFM sensitivity (see SI).

**Mechanically-tunable quantum point contact**

The low friction between Au-hBN enables studies of vdW quantum devices in which reconfigurable gold gates are used to mechanically-modify electron confinement. Gate-defined quantum point contacts (QPC) and quantum dots are of particular interest as they are integral for making graphene-based qubits (*15–17*) and for studying non-abelian quasiparticles (*18*, *19*).

We apply this unique confinement control capability to make a reconfigurable QPC defined by movable gold-only top gates on a hBN-encapsulated graphene device (Figure 2a). The top gates confine electrons by depleting the graphene into a band gap, thereby forming a narrow QPC constriction between two conducting regions. Although graphene lacks an intrinsic band gap, one forms in a perpendicular magnetic field at zero density due to exchange interactions (*20*). Therefore, in the quantum hall regime, we can study the edge mode transmission through the constriction at different Landau level filling factors and QPC separations by holding the dual-gated region at the charge neutrality point while sweeping the back-gate voltage (*21*, *22*).

To adjust the QPC separation, the top gates are physically moved with an AFM tip at ambient conditions, modifying the QPC confinement mechanically (Figure 2a). We then cool the sample to 1.5 K, apply a 9 T out-of-plane magnetic field, and measure the resistance versus top-gate and back-gate voltages. From these measurements we determine the QPC conductance, referred to as $G_{QPC}$, by taking the inverse of the measured resistance after subtracting a contact resistance (see SI for more details).

Figure 2b and 2c schematics show open and closed QPC configurations where the Landau level filling factors are $\nu_{bg} = -2$ and $\nu_d = 0$; $\nu_{bg}$ and $\nu_d$ are the back-gate and dual-gated regions, respectively. In the open configuration, the QPC separation is large and edge modes transmit across the device unimpeded resulting in a $G_{QPC}$ of 2 $e^2/h$ (Figure 2b). In contrast, for the same filling factors in the closed configuration, the QPC separation is small enough for the counter-propagating edge modes to tunnel couple and backscatter, decreasing $G_{QPC} \leq 1\ e^2/h$ (Figure 2c).

Figure 2d shows QPC conductance linetraces with $\nu_d = 0$ for four separations: 1110 nm, 170 nm, 80 nm, and 10 nm. It is apparent that physically narrowing the QPC generally decreases $G_{QPC}$ at a given back-gate voltage. Focusing on filling factor $\nu_{bg} = -2$ (vertical dashed line at -1.65 V), a 2 $e^2/h$ plateau is observed for the 1110 nm separation; this corresponds to the edge modes in the back-gated region transmitting across the device unimpeded. We refer to this as the open configuration, equivalent to Figure 2b. At the same back-gate voltage, narrowing the QPC separation to 170 nm decreases $G_{QPC}$ to between 1 and 2 $e^2/h$, indicating partial reflection of one edge mode. Further narrowing to 80 nm results in a 1 $e^2/h$ plateau. This surviving quantized plateau is explained by the spatial separation of the edge modes. The innermost counterpropagating modes are close enough to completely backscatter via tunneling, while the outer modes are still too far apart to couple and instead transmit through the QPC unimpeded. Narrowing the QPC separation further to 10 nm results in partial reflection of the remaining edge mode such that $G_{QPC} < 1\ e^2/h$ (as illustrated in Figure 2c). Similarly, for the 1 $e^2/h$ plateau in the open configuration, we are able to reduce $G_{QPC}$ to 0.14 $e^2/h$ by narrowing the QPC separation, demonstrating our ability to mechanically pinch-off the conductance.

The full dual-gate dependence of the QPC for different separations shows how the entire sequence of conductance plateaus is modified by the gate separation (Figure 2e-h). Mechanical tuning of gates offers an unprecedented level of control, as confinement geometry and physical position of the QPC can be modified independent of gate voltages. Another practical advantage is that we are able to achieve an estimated gate separation of 10 nm, which would be challenging to produce using conventional fabrication methods. Such an approach will be highly useful for tuning the properties of gate-defined quantum dots and QPCs in vdW heterostructures.

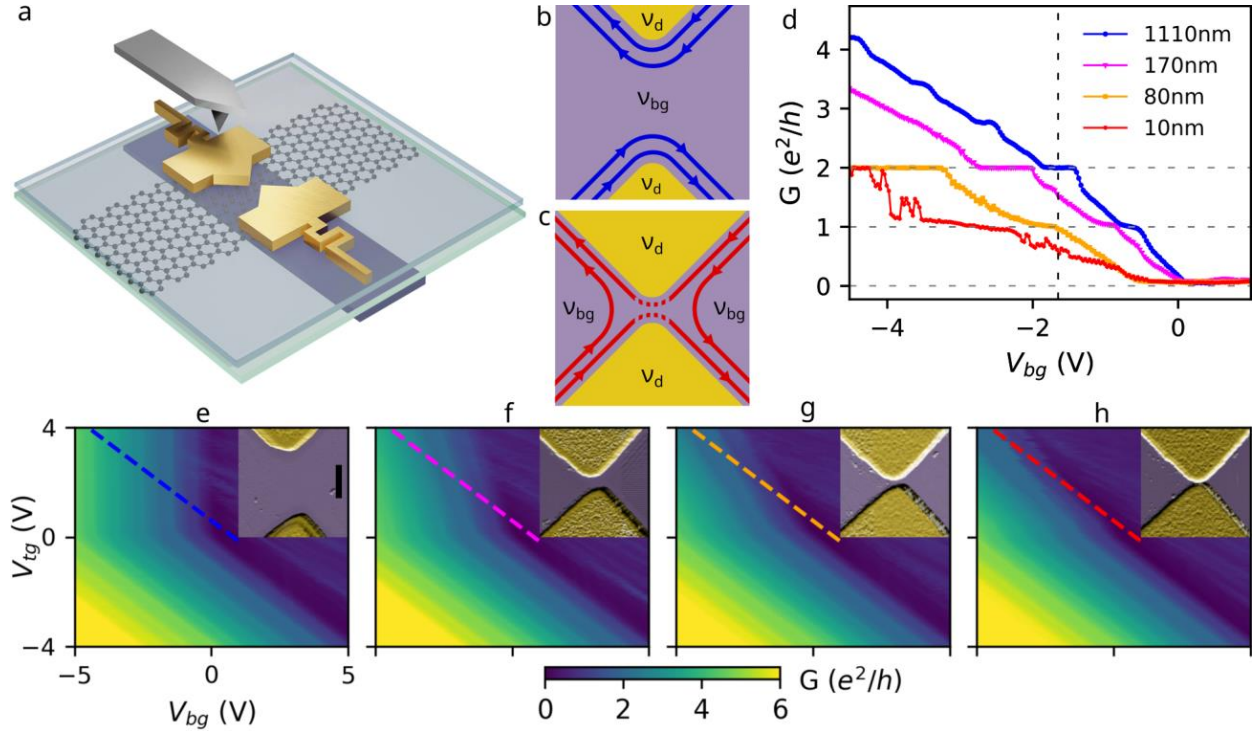

**Figure 2: Measurement of a mechanically-reconfigurable quantum point contact device**
**(a)** Schematic of an hBN-encapsulated graphene device with a local graphite back-gate and flexible serpentine leads connected to the movable QPC top gates (metal contacts to the graphene and graphite not shown). **(b&c)** QPC edge mode schematic for $\nu_d = 0$ and $\nu_{bg} = -2$. **(b)** Open configuration: all edge modes are completely transmitted, as in the 1110 nm separation. **(c)** Closed configuration: the innermost edge mode is completely backscattered while the outer edge mode is partially backscattered (indicated by the dotted lines), as in the 10 nm separation. **(d)** Line cuts of full 2D conductance color plots taken at 9 T and 1.5 K along $\nu_d = 0$ for each of the four separations. **(e,f,g,h)** Conductance color plots versus graphite back-gate and QPC top-gate voltages at separations of 1110 nm, 170 nm, 80 nm, and 10 nm, respectively. Dashed lines correspond to $\nu_d = 0$ linecuts presented in (d). Insets are false color AFM amplitude images of QPC gates. Scale bar in (e) is 500 nm and applies to all AFM images.

**In-situ heterostructure and cryogenic manipulation**

An exciting aspect of the low friction between gold and hBN is the potential for true *in-situ* manipulation of a device's atomic structure. Here, *in-situ* means simultaneous manipulation and measurement under the extreme conditions often required for quantum experiments, such as cryogenic temperatures, high magnetic fields, and high vacuum. Of these conditions, cryogenic manipulation presents the biggest challenge because friction typically increases significantly at low temperatures due to reduced thermal vibrations and the freezing of gas within the cryogenic vacuum space (*23*).

To advance vdW manipulation beyond pushing individual flakes at ambient conditions, we aim to achieve deterministic lateral motion of flakes and even whole heterostructures by creating a better mechanical interface. The conventional method for vdW flake manipulation, as demonstrated in Figures 1 & 2 and prior works (*9–12*) uses a sharp AFM tip to push a flake from the side and then reimages the flake position using the same tip. This, however, does not result in deterministic, one-to-one motion of the flake and makes certain manipulation applications such as scanning entirely infeasible. Likewise, this manipulation technique is not well-suited to overcome the high friction forces at cryogenic temperatures and for large area structures due to the small contact area between the sharp tip and the flake. In fact, we encounter the limits of this style of motion with very large gold contacts in our QPC devices (area > 35 µm$^2$), which are cut by the AFM tip as it pushes laterally. To address these issues, we have created metal handles which interface the AFM tip to a vdW heterostructure. The handle grips the vdW flakes/heterostructures by overlapping the flake edges, so that it conforms to the flake and distributes force along the flake's edges. Using a flattened AFM tip, we interface with the handle by press-fitting into a donut-shaped hole in the handle, which can be removed simply by lifting the AFM tip. The press-fit deforms the metal to match the tip shape and provides increased grip for deterministic 2D manipulations.

The interface between the AFM tip and metal handle is strong enough to enable the deterministic sliding of gold on hBN at cryogenic temperatures (T = 7.6 K, see Figure 3a and video in SI). Despite the increased friction at cryogenic temperatures, as evidenced by visible tip flex in the video, we observe that the hBN surface is left pristine and undamaged after more than 100 motions at a speed of 30 µm/s (Figure 3b). See SI for more details on the *in-situ* sliding technique and manipulation setup. Cold sliding can enable a variety of experiments including reconfigurable vdW heterostructures at cryogenic temperatures, which allows for rapid, continuous measurements with respect to physically reconfigurable parameters.

While mechanical linkage and motion is useful in its own right, electrical contact to moveable structures is also critical to perform many experiments. To this end, we fabricate flexible serpentine-shaped electrodes connected to our donut handles, shown in Figure 3c. We are able to oscillate these electrodes at 10 Hz with an amplitude of 2 µm by actuating with an AFM tip and the same donut interface described above (video available online). We find these accordion geometries with a wire cross section of 1 µm thick and 1 µm wide are able to stretch over 10 µm before breaking.

By combining mechanical motion and flexible electrodes, we create a sliding scanning top-gate, shown in Figure 3d-f. At room temperature, we raster a gold top-gate by sliding it over an encapsulated graphene-hBN device, modulating the device resistance by changing the overlap between the graphene and top-gate. The resistance versus gate position is plotted in Figure 3f, which can be interpreted as a coarse image of our graphene device convolved with the geometry of the sliding gate. This constitutes a new mechanism for scanning probe microscopy (*24*), where a gate is in direct atomic contact with the sample, obviating the need for the feedback control of the probe-sample distance that is typical with scanning probes.

Taking advantage of the low-friction of both graphene-hBN and Au-hBN interfaces, we apply our technique to make a slidable, contacted vdW heterostructure, shown optically and schematically in Figure 3g,h. Here an entire graphene device is translated over an hBN substrate, actuated via a metal handle. The graphene, edge-contact electrodes, and top hBN all slide as a single unit, and allow for continuous measurement of the graphene as it is moved. Figure 3i shows the change in graphene resistance as it is translated back and forth from the position where it was initially transferred. As the graphene slides over 1.2 µm, we observe a reproducible modulation of the device resistance corresponding to a maximum change of 10-15 Ω. One explanation for this effect is that the graphene acts as a local charge sensor by varying in resistance as it moves through potential inhomogeneity across the hBN surface (*25*). Another effect which can arise in this device geometry is strain in the graphene or at the graphene-gold interface that develops in response to friction forces during the motion. These effects will be isolated and explored in future studies.

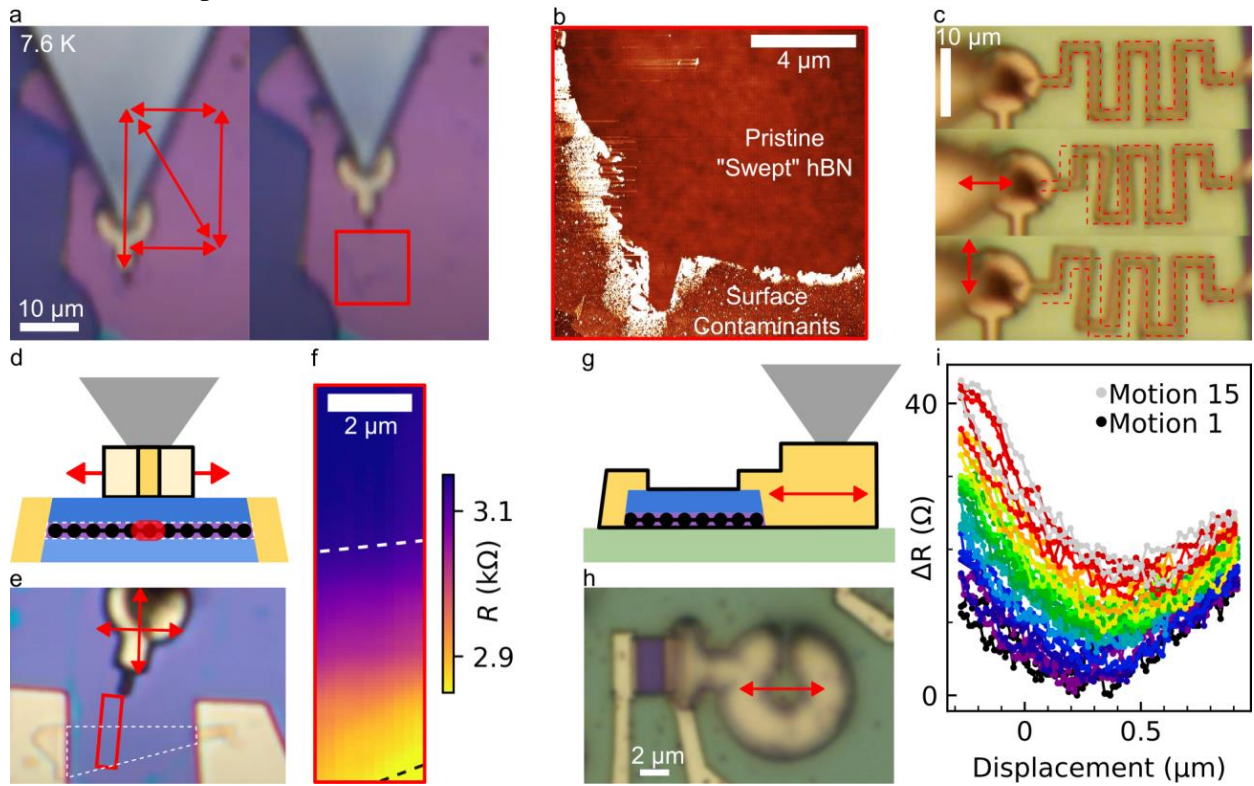

**Figure 3:** *In-situ* **mechanically-reconfigurable devices**

**(a)** Optical images showing gold sliding on hBN at 7.6 K. Arrows denote the range and direction of motions. Video of oscillating motion available online. **(b)** AFM image of hBN surface after cryogenic scanning motions from (a) showing the hBN surface is left undamaged with only swept up surface contaminants. AFM area is the solid red boxed area in (a). **(c)** Optical images of gold serpentine electrodes showing ~2 µm longitudinal and transverse motions. Red dotted lines outline the initial position. Video of oscillating motion available online. **(d)** Side profile schematic of a sliding top-gate hBN-encapsulated graphene device with AFM tip. Top-gate slides over stationary graphene to change local gating and device resistance. **(e)** Top-down optical image of the same device. Graphene outlined with a dotted white line. Red rectangle is 2.4 x 9.5 µm. **(f)** Graphene resistance versus top-gate position. Scanning range shown as the red rectangle in the optical image. Dashed lines indicate graphene edge. **(g)** Side profile schematic of a slidable graphene-hBN device transferred on a stationary hBN substrate with AFM tip. The slidable features in the schematic are outlined in black. **(h)** Top-down optical image of the same device. **(i)** 2-probe resistance of the slidable graphene device versus sliding position. 0 µm corresponds to the initial, transferred position. The continuous increase in resistance over subsequent motions is likely due to photodoping of the graphene from the white light source used for imaging (see SI for videos).

**Discussion**

The ability to move both metal and vdW layers within a device offers an unprecedented level of control and flexibility in both device function and experiment design. The mechanically-reconfigurable devices we demonstrate enable experimental studies where structure and geometry are continuously tunable parameters. This allows for dense sampling of the device and heterostructure parameter space while keeping local disorder constant, something that is impossible to achieve with the conventional approach of fabricating multiple devices. Reconfiguration by sliding makes possible the modification of quantum confinement via moveable gate electrodes, as well as the continuous tuning of lattice interfaces in vdW moire heterostructures. Our demonstration of deterministic *in-situ* sliding also introduces the possibility of dynamic structural studies, where time-varying modulations of the device geometry and interfacial moires induce electronic effects such as topological charge pumping (*26–28*). Lastly, the proof-of-principle sliding scanning probe experiments show a new approach to spatial mapping of local material properties at the extreme limit of proximity, i.e. direct atomic contact, as well as with the full flexibility of planar nanofabrication.

**References**
1. *Microsystem Design* (Kluwer Academic Publishers, Boston, 2002; http://link.springer.com/10.1007/b117574).
2. M. Dienwiebel, G. S. Verhoeven, N. Pradeep, J. W. M. Frenken, J. A. Heimberg, H. W. Zandbergen, Superlubricity of Graphite. *Phys. Rev. Lett.* **92**, 126101 (2004).
3. O. Hod, E. Meyer, Q. Zheng, M. Urbakh, Structural superlubricity and ultralow friction across the length scales. *Nature*. **563**, 485–492 (2018).

4. E. Koren, E. Lörtscher, C. Rawlings, A. W. Knoll, U. Duerig, Adhesion and friction in mesoscopic graphite contacts. *Science*. **348**, 679–683 (2015).
5. Y. Song, D. Mandelli, O. Hod, M. Urbakh, M. Ma, Q. Zheng, Robust microscale superlubricity in graphite/hexagonal boron nitride layered heterojunctions. *Nat. Mater.* **17**, 894–899 (2018).
6. H. Rokni, W. Lu, Direct measurements of interfacial adhesion in 2D materials and van der Waals heterostructures in ambient air. *Nat. Commun.* **11**, 5607 (2020).
7. Z. Liu, J. Yang, F. Grey, J. Z. Liu, Y. Liu, Y. Wang, Y. Yang, Y. Cheng, Q. Zheng, Observation of Microscale Superlubricity in Graphite. *Phys. Rev. Lett.* **108**, 205503 (2012).
8. M. Liao, P. Nicolini, L. Du, J. Yuan, S. Wang, H. Yu, J. Tang, P. Cheng, K. Watanabe, T. Taniguchi, L. Gu, V. E. P. Claerbout, A. Silva, D. Kramer, T. Polcar, R. Yang, D. Shi, G. Zhang, Ultra-low friction and edge-pinning effect in large-lattice-mismatch van der Waals heterostructures. *Nat. Mater.* **21**, 47–53 (2022).
9. T. Chari, R. Ribeiro-Palau, C. R. Dean, K. Shepard, Resistivity of Rotated Graphite–Graphene Contacts. *Nano Lett.* **16**, 4477–4482 (2016).
10. E. Koren, I. Leven, E. Lörtscher, A. Knoll, O. Hod, U. Duerig, Coherent commensurate electronic states at the interface between misoriented graphene layers. *Nat. Nanotechnol.* **11**, 752–757 (2016).
11. R. Ribeiro-Palau, C. Zhang, K. Watanabe, T. Taniguchi, J. Hone, C. R. Dean, Twistable electronics with dynamically rotatable heterostructures. *Science*. **361**, 690–693 (2018).
12. N. R. Finney, M. Yankowitz, L. Muraleetharan, K. Watanabe, T. Taniguchi, C. R. Dean, J. Hone, Tunable crystal symmetry in graphene–boron nitride heterostructures with coexisting moiré superlattices. *Nat. Nanotechnol.* **14**, 1029–1034 (2019).
13. D. Dietzel, A. S. de Wijn, M. Vorholzer, A. Schirmeisen, Friction fluctuations of gold nanoparticles in the superlubric regime. *Nanotechnology*. **29**, 155702 (2018).
14. D. Dietzel, M. Feldmann, U. D. Schwarz, H. Fuchs, A. Schirmeisen, Scaling Laws of Structural Lubricity. *Phys. Rev. Lett.* **111**, 235502 (2013).
15. L. Banszerus, A. Rothstein, E. Icking, S. Möller, K. Watanabe, T. Taniguchi, C. Stampfer, C. Volk, Tunable interdot coupling in few-electron bilayer graphene double quantum dots. *Appl. Phys. Lett.* **118**, 103101 (2021).
16. L. A. Cohen, N. L. Samuelson, T. Wang, K. Klocke, C. C. Reeves, T. Taniguchi, K. Watanabe, S. Vijay, M. P. Zaletel, A. F. Young, Tunable fractional quantum Hall point contacts in graphene via local anodic oxidation of graphite gates (2022), (available at http://arxiv.org/abs/2204.10296).
17. H. Overweg, H. Eggimann, X. Chen, S. Slizovskiy, M. Eich, R. Pisoni, Y. Lee, P. Rickhaus, K. Watanabe, T. Taniguchi, V. Fal'ko, T. Ihn, K. Ensslin, Electrostatically Induced Quantum Point Contacts in Bilayer Graphene. *Nano Lett.* **18**, 553–559 (2018).
18. C. Déprez, L. Veyrat, H. Vignaud, G. Nayak, K. Watanabe, T. Taniguchi, F. Gay, H. Sellier, B. Sacépé, A tunable Fabry–Pérot quantum Hall interferometer in graphene. *Nat. Nanotechnol.* **16**, 555–562 (2021).
19. J. Nakamura, S. Liang, G. C. Gardner, M. J. Manfra, Direct observation of anyonic braiding statistics. *Nat. Phys.* **16**, 931–936 (2020).
20. A. F. Young, C. R. Dean, L. Wang, H. Ren, P. Cadden-Zimansky, K. Watanabe, T. Taniguchi, J. Hone, K. L. Shepard, P. Kim, Spin and valley quantum Hall ferromagnetism in graphene. *Nat. Phys.* **8**, 550–556 (2012).
21. K. Zimmermann, A. Jordan, F. Gay, K. Watanabe, T. Taniguchi, Z. Han, V. Bouchiat, H. Sellier, B. Sacépé, Tunable transmission of quantum Hall edge channels with full degeneracy lifting in split-gated graphene devices. *Nat. Commun.* **8**, 14983 (2017).
22. Y. Ronen, T. Werkmeister, D. Haie Najafabadi, A. T. Pierce, L. E. Anderson, Y. J. Shin, S. Y. Lee, Y. H. Lee, B. Johnson, K. Watanabe, T. Taniguchi, A. Yacoby, P. Kim, Aharonov–Bohm effect in graphene-based Fabry–Pérot quantum Hall interferometers. *Nat.*

23. X. Zhao, M. Hamilton, W. G. Sawyer, S. S. Perry, Thermally Activated Friction. *Tribol. Lett.* **27**, 113–117 (2007).
24. M. A. Topinka, B. J. LeRoy, S. E. J. Shaw, E. J. Heller, R. M. Westervelt, K. D. Maranowski, A. C. Gossard, Imaging Coherent Electron Flow from a Quantum Point Contact. *Science*. **289**, 2323–2326 (2000).
25. J. Xue, J. Sanchez-Yamagishi, D. Bulmash, P. Jacquod, A. Deshpande, K. Watanabe, T. Taniguchi, P. Jarillo-Herrero, B. J. LeRoy, Scanning tunnelling microscopy and spectroscopy of ultra-flat graphene on hexagonal boron nitride. *Nat. Mater.* **10**, 282–285 (2011).
26. M. Fujimoto, H. Koschke, M. Koshino, Topological charge pumping by a sliding moiré pattern. *Phys. Rev. B*. **101**, 041112 (2020).
27. Y. Zhang, Y. Gao, D. Xiao, Topological charge pumping in twisted bilayer graphene. *Phys. Rev. B*. **101**, 041410 (2020).
28. Y. Su, S.-Z. Lin, Topological sliding moiré heterostructure. *Phys. Rev. B*. **101**, 041113 (2020).
29. F. Pizzocchero, L. Gammelgaard, B. S. Jessen, J. M. Caridad, L. Wang, J. Hone, P. Bøggild, T. J. Booth, The hot pick-up technique for batch assembly of van der Waals heterostructures. *Nat. Commun.* **7**, 11894 (2016).
30. D. G. Purdie, N. M. Pugno, T. Taniguchi, K. Watanabe, A. C. Ferrari, A. Lombardo, Cleaning interfaces in layered materials heterostructures. *Nat. Commun.* **9**, 5387 (2018).

**Acknowledgments**

The authors acknowledge the use of facilities and instrumentation at the Integrated Nanosystems Research Facility (INRF), in the Samueli School of Engineering at the University of California Irvine, and at the UC Irvine Materials Research Institute (IMRI), which is supported in part by the NSF MRSEC through the UC Irvine Center for Complex and Active Materials. The authors also acknowledge the use of the UCI Laser Spectroscopy Lab. The authors thank L. Jauregui and M. Yankowitz for productive discussions, as well as the technical assistance of Q. Lin, R. Chang, M. Kebali, J. Hes, and D. Fishman.

**Funding:** NSF Career Award 2046849

**Author contributions:**
Investigation: AZB, IS, YY, AHBA, JDSY
Sample Preparation: AZB, IS, AHBA
Supervision: JDSY
Writing & Reviewing: AZB, IS, JDSY, AHBA

**Competing interests:** Authors declare that they have no competing interests

**Data and materials availability:** Data available upon request.

# Supplemental Information

## Mechanically-reconfigurable van der Waals devices via low-friction gold sliding

### General fabrication techniques

The following techniques are universal to our device fabrication except when specified otherwise.

**Lithography**

All lithography performed is electron beam lithography (EBL) using a PMMA resist. We use PMMA 950 A5 spun at 2000 rpm for 2 minutes resulting in a ~500 nm thick layer for depositions less than 300 nm in thickness and for etch masks. EBL patterns are written at 1.6 nA or 3.2 nA with 30 kV excitation. The PMMA is developed for 3 minutes in a cold mixture of 3:1 IPA:water.

**1D-edge-contacts**

1D-edge-contacts for our graphene encapsulated in hexagonal boron nitride (hBN) devices are written with EBL and developed before reactive ion etching (RIE) with 10 sccm of $SF_6$, 2 sccm of $O_2$, 30 W of RF power, at 100 mTorr for 30 s (*29*). Then 3 nm of Cr and ~100 nm of Au are deposited at 1 Å/s in an electron beam metal vapor deposition system. Liftoff is performed by soaking the sample in acetone for 1-2 hours and agitating with a pipette.

**Dry transfers**

Stacks are assembled using stamps consisting of PC film on a PDMS square on a glass slide (*30*).

### Friction measurements

**Fabrication for gold-hBN friction measurements**

Gold-only squares for friction measurements are fabricated by spinning PMMA on a silicon chip of exfoliated hBN, writing squares using EBL, depositing 170 nm of gold, and lifting off in acetone.

**AFM lateral friction measurements**

To measure lateral friction we use the lithography mode of a Park Systems NX10 AFM and a Budget Sensors Tap300Al-G tip to manipulate gold on hBN in ambient conditions. First the top surface of the gold squares is measured using lithography set point mode with 100 nN of downward force and a dwell time of 8 s. The tip is then moved next to the gold square, lowered 100 nm below the top surface of the gold, and moved laterally, transverse to the AFM cantilever, and into the gold square at 10 nm/s. The lateral deflection voltage is recorded throughout the motion. Measurements were also made at 1 nm/s and 100 nm/s, and no dependence on speed was observed in this range. Care is taken for the motion to be through the center of mass of the gold in order to avoid rotation, and motions with rotation are excluded from our analysis. The squares are also oriented so their edge is perpendicular to the direction of motion. Note, that no permanent deformation of the gold is observed in AFM images taken after manipulations.

The regions before contact, during static friction, and during kinetic friction are identified to extract the average baseline voltage, average kinetic friction, and the peak static friction value respectively. We include some example linetraces to demonstrate the scales, and overall appearance of typical linetraces.

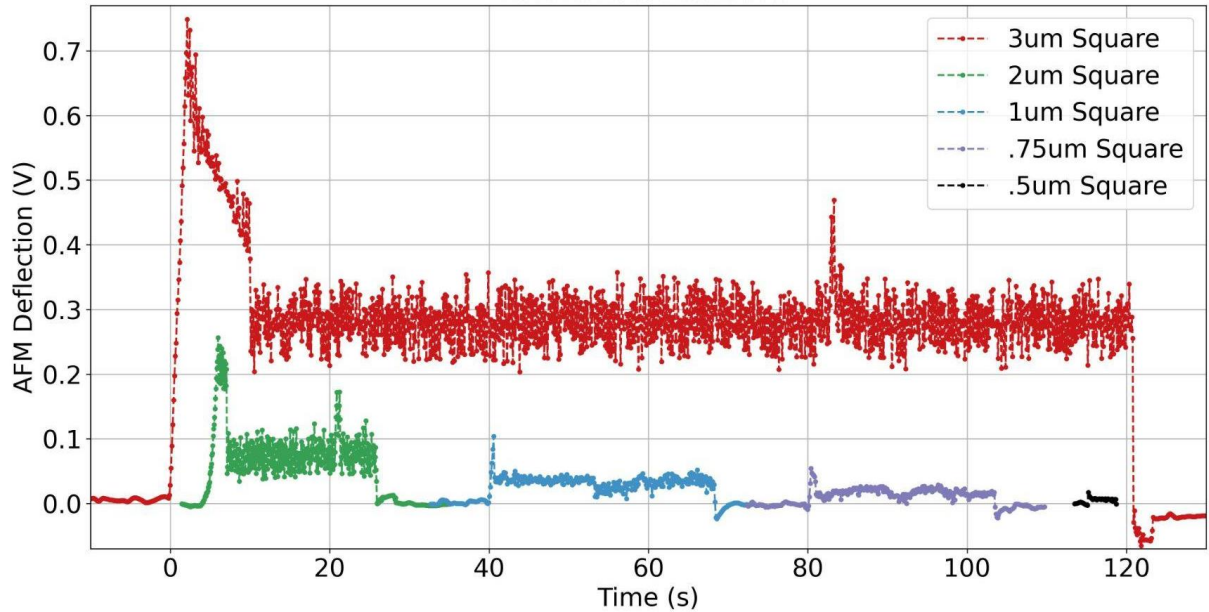

**Figure S1: Example AFM deflection linetraces for various sizes of gold squares on hBN.**

To measure force using the AFM, we assume a linear model where deflection voltage is related to the force on the cantilever as $F_{tip} = V_{tip} \, k / S$, where $V_{tip}$ is the lateral deflection voltage of the AFM tip, $k$ is the tip spring constant, and $S$ is the AFM sensitivity.

The cantilever is simulated in COMSOL Multiphysics to determine its spring constant, referencing optical and SEM images for its dimensions (6). To simulate the spring constant a lateral force is applied 100 nm above the apex of the tip and the displacement at this position is determined, resulting in a force per distance displaced of $k = 250 \pm 10$ N/m for our tip. The simulations are parameterized with respect to the cantilever dimensions in order to estimate an uncertainty for the lateral spring constant. The material used is single crystal anisotropic silicon with the top of the cantilever as the <100> plane and the cantilever pointing in the <110> direction.

To determine the sensitivity, $S$, which is the ratio of the lateral voltage deflection to tip displacement, we take the slope of the static region of the deflection linetrace, when the tip deflects before the gold starts moving. This assumes that the lateral displacement of the AFM piezo stage is equal to the tip deflection during the static portion of the pushing. However, we expect the tip deflection to be less than the stage displacement due to other effects such as the elastic deformation of the gold, or due to the tip slipping as it comes into contact with the gold edge. Such effects are evidenced by non-linearity in the slope of the static region and by the variable sensitivities we observe for smaller squares. The net effect is that our approach will overestimate the displacement of the cantilever, which would result in our extracted forces providing an upper bound on the friction force. Due to the variable sensitivity we observe for smaller squares (1 µm$^2$ and smaller), we only use the 4 and 9 µm$^2$ linetraces to calculate an average lateral sensitivity of $S = 27 \pm 8$ mV/nm.

**hBN-graphite friction**

We measure several flakes of hBN on graphite and see that while the friction for 6 µm$^2$ flake matches very well for the same sized gold square, larger hBN flakes on graphite exhibit similarly low friction and hence indicate a sub-linear scaling with area, unlike the polycrystalline gold on hBN (Figure S2). This is expected since the hBN flakes are single crystal and exhibit superlubricity (*5*, *8*). The hBN flakes measured are 20-50 nm tall, much shorter than the 170 nm tall gold squares, so to manipulate the thin hBN flakes, the height of the underlying graphite substrate is measured with 100 nN downward force. This sets the AFM tip height for the lateral motion. We find that using this height setting does not result in the tip touching the graphite substrate during manipulation.

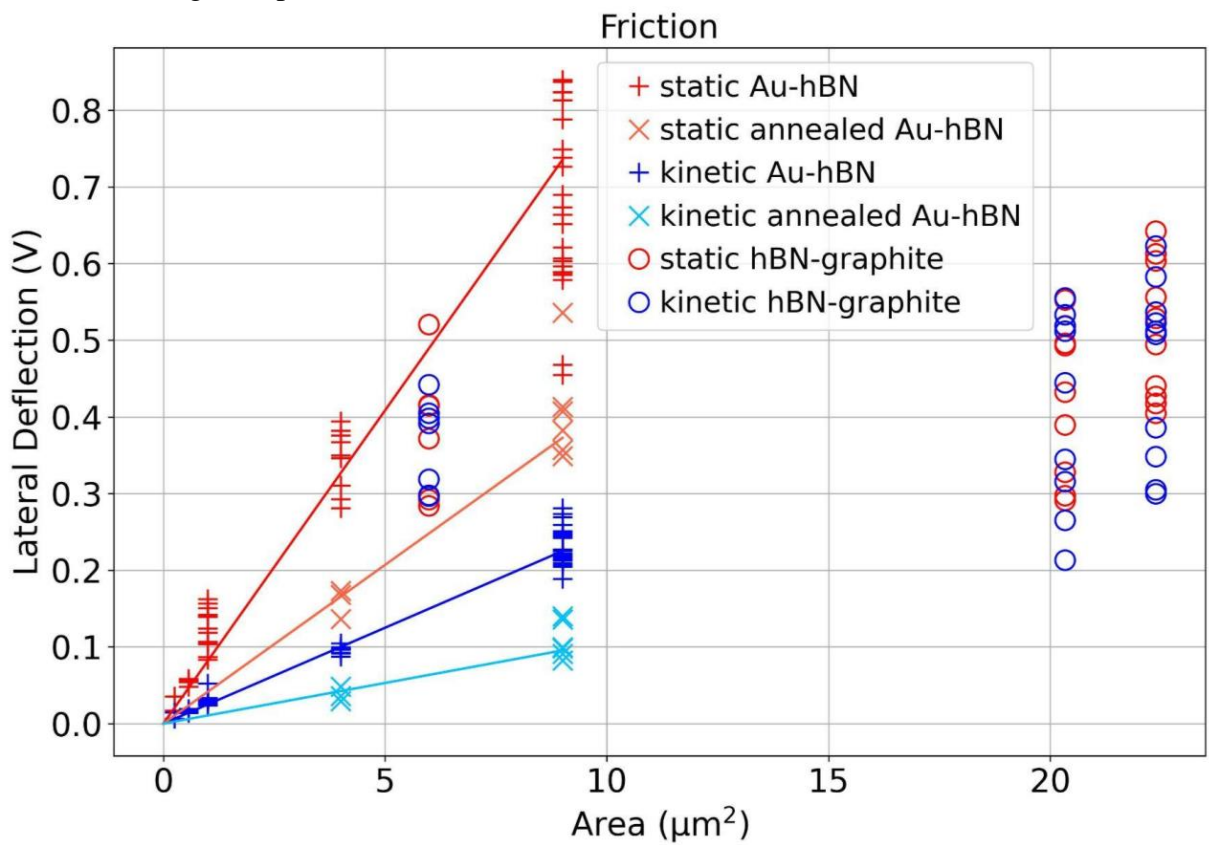

**Figure S2**: **All static and kinetic deflection values for gold-on-hBN and hBN-on-graphite extracted from linetraces.**

Friction forces per unit area for Au-hBN are: static 0.762 ± 0.232 MPa, static annealed 0.386 ± 0.117 MPa, kinetic 0.233 ± 0.07 MPa, kinetic annealed 0.099 ± 0.031 MPa. Fits are through zero and only include 4 and 9 µm$^2$ data points.

**Annealing gold on hBN**

To test the effect of heat annealing, the gold squares on hBN are vacuum annealed for 30 minutes at 350 C in a tube furnace. Measuring their friction again, we see that the tip deflections decreased by ~50%. AFM imaging shows that the grain sizes increased from ~20 nm to 80 nm in diameter. This qualitative behavior is expected for polycrystalline materials, but it may also be the result of trapped contaminants escaping from the gold-hBN interface.

**Cr-Au friction on hBN**

When moving small pieces of 3nm Cr + 100 nm Au on hBN, we have found their static and kinetic deflection voltages to be roughly an order of magnitude higher than the same sized annealed gold-only square on hBN.

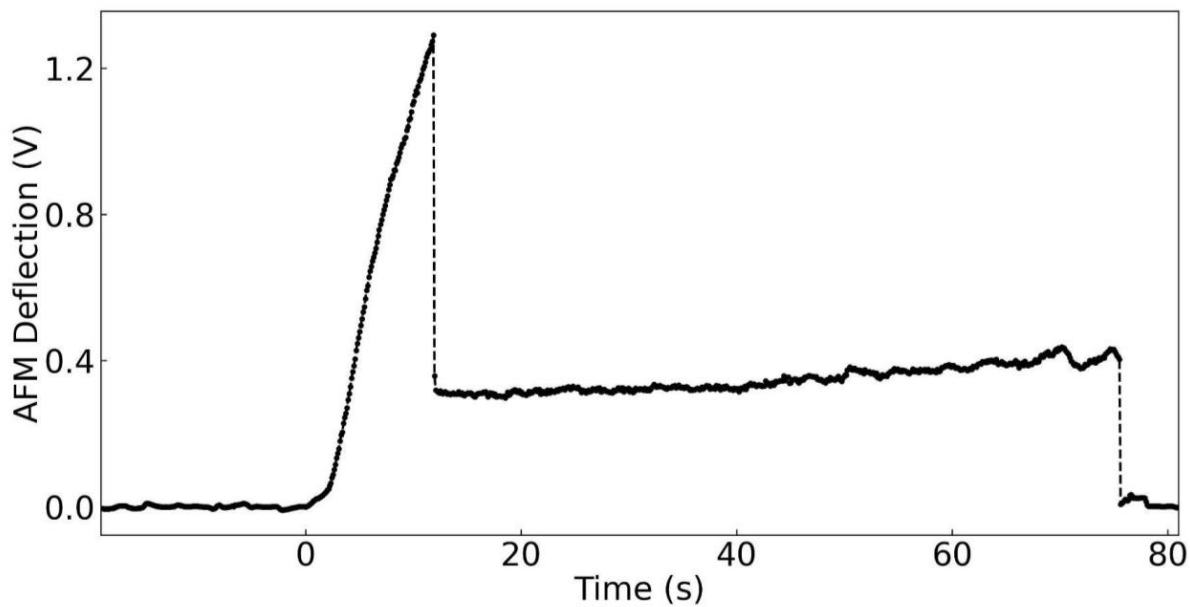

**Figure S3: Voltage deflection for 2 µm x 2 µm, 100 nm thick gold + 3 nm Cr sticking layer square on hBN**

## Quantum point contact device

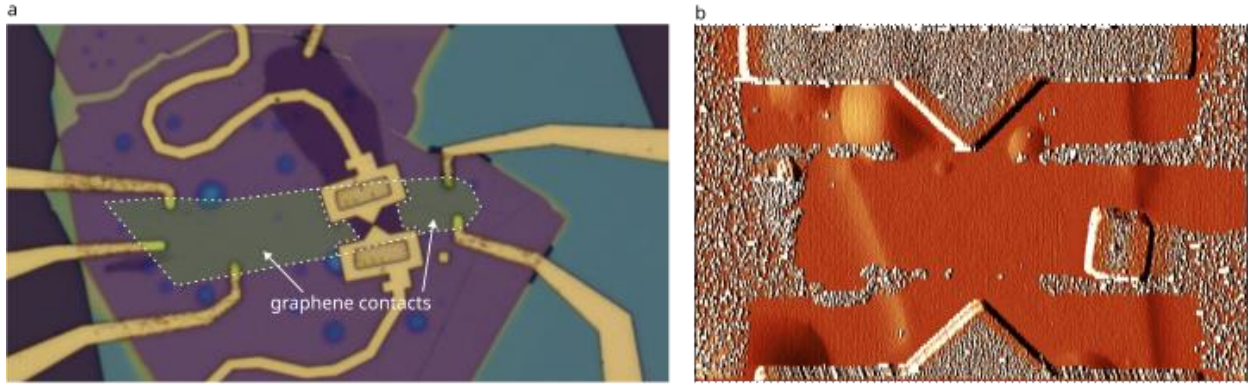

**Figure S4: Images of mechanically reconfigurable quantum point contact device**
**(a)** 100x optical image of completed QPC device after finishing first set of gate manipulations to achieve a desired gate separation. Graphene flake is outlined and graphene contact regions highlighted in green reside in regions where the carrier density is determined only by the silicon back-gate. **(b)** AFM amplitude data taken zoomed in on the QPC gates showing the cleaning process. A gold square is swept back and forth between the QPC gates to remove surface contaminants from the sensitive QPC region prior to positioning the QPC gates to a desired separation.

### QPC device fabrication

Our reconfigurable QPC device consists of a graphene strip encapsulated in hBN, with a local graphite back-gate on an Si/SiO$_2$ substrate. 1D-edge-contacts are added to the graphene and graphite back-gate. The moveable top gates are 170 nm thick gold-only deposited with long, flexible electrodes to allow the gates to be moved while maintaining electrical contact. The Au-hBN interface friction is relatively large for the top gates and flexible electrodes due to their large surface area. To avoid cutting the gates while attempting motions we cross-link 500 nm tall PMMA rectangles onto the gates. These tall features provide more surface area for the AFM tip to push into in order to move the top gates. Although they still deform from manipulations they serve as a sacrificial handle. We dose the PMMA with 15000 µC/µm$^2$ at 30 kV and 3.2nA in order to cross-link it. A small, isolated square of gold is also deposited onto the hBN (at the same time as the gold-only top gates), so that contaminants can be swept from the hBN surface, seen in Fig S4b

The graphite back-gate for our device is smaller than the graphene. This is so the 1D-edge-contacts to the graphene don't short to the back-gate. We refer to regions of the graphene which are not gated by the graphite as the graphene contacts, and they are doped to a higher carrier density by the silicon back-gate.

**QPC measurement details**

The resistance data was taken in a four-probe configuration by measuring the current at the drain electrode via a Femto current preamplifier (1E6 V/A gain) and the longitudinal voltage drop with a SR830 lockin amplifier using low frequency lockin techniques.

Before performing the 2D gate sweeps, we hole-dope the graphene contacts by biasing the silicon back-gate with -45 V corresponding to a filling factor of ~7 at 9 T (nominal 285 nm $SiO_2$ thickness). For this reason, when both the dual-gated region and back-gated region are electron doped the device is unable to effectively transmit quantum hall edge modes across the resulting pn-junction, and appears insulating as evident in the upper right corner of each conductance color plot in Figure 2e-h of the main text.

**QPC contact resistance**

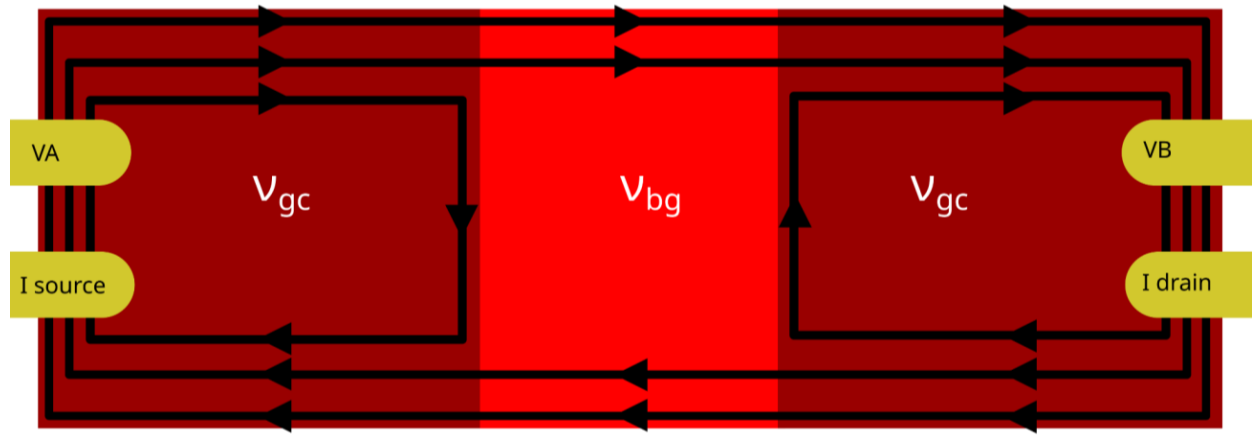

**Figure S5: Three hole-doped quantum hall regions in series**.
Electrodes in yellow are set up in a four-probe configuration, identical to the measurements for the QPC device. $\nu_{gc}$ is the filling factor of the graphene contacts and $\nu_{bg}$ is the filling factor of the graphite back-gated region where $|\nu_{gc}| > |\nu_{bg}|$.

Due to the graphene contacts being in series with the QPC constriction, the measured resistance is a sum of the QPC constriction resistance and the graphene contacts resistance $R_C$. This constant resistance $R_C$ is subtracted from the measured value to calculate $G_{QPC}$ which is presented in Figure 2 of the main text. $R_C$ was determined by fitting the value of the widest plateau in Figure 2d to a conductance of 2 $e^2/h$, which we ascribe as an effective $\nu_{qpc} = -2$ at the QPC constriction point. Note that the graphene contacts support their own quantum hall edge states, generally at a different filling factor than the back-gated and dual-gated regions. In a four-probe longitudinal resistance measurement between a network of different quantum hall regions with different filling factors, one can measure effective positive and negative contact resistances as outlined below. For the open configuration, we measured a net positive contribution to the contact resistance and for the closed configurations we measured a net negative contribution. The contact resistance values used are $R_C$ =(+1580, -3000, -2920, -2800 Ω) for the QPC separations of 1110 nm, 170 nm, 120 nm, and 10 nm respectively.

A positive contribution to $R_C$ can be understood due to disorder causing forward moving edge modes to backscatter to backward moving edge modes and vice versa. Bubbles formed between the layers during the stacking process are a source of disorder in our device and can be seen optically or in our AFM images in the graphene contact regions.

As an example of how one can measure a negative contribution to $R_C$ with our contact configuration, consider the scenario depicted in Figure S4 with three quantum hall regions in series with filling factors from left to right $\nu_{gc}$, $\nu_{bg}$, and $\nu_{gc}$ where $\nu_{gc}$ is the landau level filling factor of the graphene contact regions. Assuming both $\nu_{gc}$ and $\nu_{bg}$ are negative (as is the case for operating the QPC), the magnitude of $|\nu_{gc}| > |\nu_{bg}|$, and full equilibration of the quantum hall edge modes, one can derive the following equations: defining $R_{AB} = V_{AB} / I_{drain}$, then $R_{AB} = R_{bg} - R_{gc}$ where $R_{bg} = h / (\nu_{bg} e^2)$ and $R_{gc} = h / (\nu_{gc} e^2)$.

## *In-situ* sliding
### Metal handle and flexible serpentine electrode fabrication

To fabricate the "thick" metal handles and flexible serpentine electrodes for *in-situ* sliding, much thicker PMMA than normal is used: 950 A11 spun at 4000 rpm for 5 s, then 2750 rpm for 2 minutes. This produces a film of PMMA that is ~2.25 µm thick. EBL is performed using our standard parameters described in the general fabrication technique section. The metal deposition consists of 80 nm Au, 10 nm Cr, 1 µm Cu, 10 nm Cr, and 50 nm Au, all at 1 Å/s except for the copper which is deposited at 3 Å/s. The bottom layer of gold is for low friction with the hBN substrate, the top gold protects against oxidation, and the chromium acts as a sticking layer between the gold and copper.

1 µm and 500 nm flexible serpentine electrode widths were tested and we find that 500 nm width electrodes are more flexible, but also more fragile. Both work well for manipulation but we prefer the 500 nm due to the decreased interface area and hence reduced friction. We have also fabricated 1 µm and 500 nm width "thin" serpentine electrodes which are 150 nm tall gold-only instead of the thick multilayer metal combination, and find that the 500 nm width electrodes are quite fragile in this case and can break more easily.

Typical metal handle donut dimensions we use have a nominal inner diameter 3 µm and an outer diameter of 9 µm. The C-shape is to help with liftoff as well as to add compliance in being stretched open upon press-fitting with the AFM tip.

### *In-situ* manipulation technique

Before performing *in-situ* manipulations using our metal handles, we first flatten a 100 N/m, 1200 MHz MicroMasch 4XC AFM tip in order to increase the tip-handle contact area. This is carried out by oscillating the tip back and forth and side to side while in contact with the $SiO_2$ of our sample until it is >3 µm wide, slightly larger than the inner diameter of the donut. We choose this tip for its high force constant and the protruding "tip view" style which makes it easier to align with the donuts.

In general, our technique for performing *in-situ* manipulations involves the sample with a metal handle donut mounted on a XYZ piezo scanner that engages into a stationary flattened AFM tip. This process is carried out by first aligning the donut with the tip and raising the piezo stage. The stage is raised until the donut contacts the tip as determined by optically observing a change in the light reflected off the cantilever (seen through a 20x objective). After the initial contact, we press-fit the tip into the donut by raising the stage further. At room temperature we engage the donut into the tip by ~1.5-2.5 µm to achieve a rigid connection. In higher friction situations, such as for larger objects or at cryogenic temperatures, engaging more than ~2.5 µm, corresponding to a larger downward force of the tip on the donut, is necessary to reduce the slippage between the tip and donut and to achieve deterministic, one-to-one sliding motions. For the sliding done at 7.6 K, we engaged ~5 µm as opposed to ~2.5 µm at room temperature to get a similar one-to-one deterministic motion.

**Room temperature manipulation setup**

Room temperature *in-situ* sliding experiments, including the scanning top gate and the sliding graphene "hockey puck" heterostructure, were performed at ambient conditions. The setup consists of an AFM tip glued to a glass slide, mounted on a coarse manual XYZ stage and the sample/chip carrier mounted on a Thorlabs 3-axis NanoMax open loop piezo stage. A Mitutoyo VMU microscope with 20x and 2x objectives is used for optical imaging. In both the room temperature and cryogenic manipulation setups, nanomanipulation is performed using the piezo positioners, which are brought into range by coarse screw-based positioners.

**Cryogenic manipulation setup**

Cryogenic manipulations are performed in a continuous flow Janis ST-500 optical cryostat, using the same microscope as for the room temperature manipulation setup. The ST-500 is cooled using a Janis helium recirculation setup and achieves a sample temperature of 7.6 K. The sample is mounted on an Attocube ANSxyz100std/LT piezo scanner (with 55 µm of XY range and 25 µm of Z range at ~4 K), and the AFM tip is mounted to a custom flexural positioner for coarse positioning at room temperature (Figure S6). Accounting for thermal contraction, the tip is coarsely positioned at room temperature so that it will be within piezo scanner range of the sample once at base temperature.

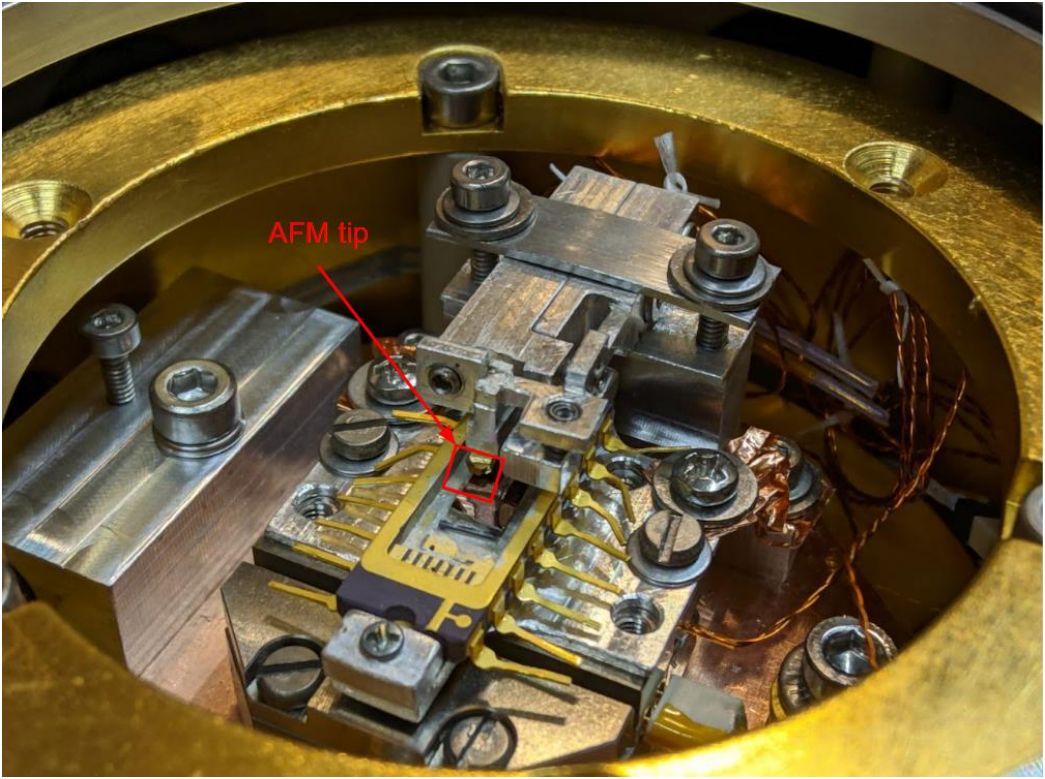

**Figure S6: custom build cryogenic manipulation setup**

**Sliding "hockey puck" heterostructure**

We fabricate our encapsulated slidable graphene "hockey puck" device by transferring hBN onto monolayer graphene, writing rectangular PMMA etch masks, and etching with $SF_6$ to leave behind rectangles of hBN on graphene. We then write and deposit 150 nm gold-only "wrap-around contacts" which provide edge contacts to the graphene and mechanical support to grip and hold the "hockey puck" heterostructures together. The "hockey pucks" are picked up using a PC/PDMS stamp and transferred onto a large bottom hBN substrate before writing and depositing thick flexible electrodes and donuts in order to electrically contact and manipulate them *in-situ*. Of the five "hockey puck" devices we attempted to transfer and electrically contact, all but one were successfully contacted.

Two-probe resistance measurements were performed by current biasing and measuring the voltage drop across the device to determine the resistance. The resistance versus displacement trace was measured by displacing in 10 nm steps and then recording the resistance after a short pause. In the main text we refer to the full forward and backward measurement sequence as a single "motion". Each motion takes about 30 seconds. A video of the motions was recorded and a light source illuminated the sample during all of the measurements. In addition, all measurements were performed with the silicon back-gate grounded.

In Figure 3d of the main text, the zero position corresponds to the initial transferred position of the "hockey puck", before any motion. After being transferred and contacted, the device was slid to the leftmost position, at about -275 nm, before the subsequent 1.2 µm back-and-forth motions.

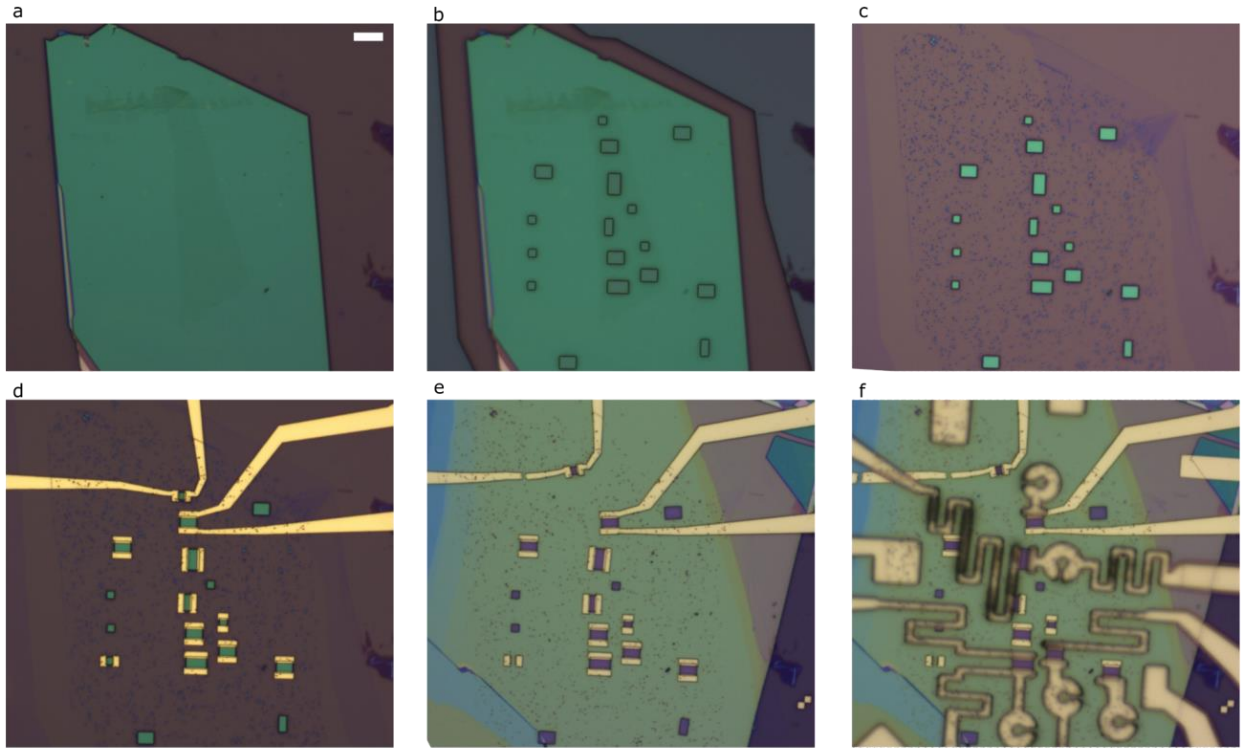

**Figure S7: Optical images of hockey puck device fabrication.**
The scale bar in (a) is 3 µm and applies to all images. **(a)** hBN transferred on top of graphene flake on $SiO_2$ substrate. **(b)** Developed PMMA etch mask written over regions of hBN with and without graphene beneath. **(c)** hBN and graphene etched using $SF_6$. **(d)** 150 nm gold-only wrap-around contacts deposited. **(e)** Hockey pucks and contacts transferred onto bottom hBN flake. **(f)** Thick metal handles and flexible serpentine electrodes deposited.

**Scanning top gate device**

The scanning top gate device was fabricated by encapsulating graphene in hBN and 1D-edge-contacting it. For this device, the "thick" metal top gate serpentine electrode structure was initially written and deposited on a separate $SiO_2$ chip and then picked up and transferred onto the completed graphene device using a PC/PDMS stamp. This electrode was transferred to a position so that one end overlapped with a previously written Cr/Au contact which provided the electrical connection to the top gate as well as a high friction anchor point (Figure S8).

Two-probe resistance measurements were performed in an identical manner to the sliding "hockey puck" device described earlier with the addition of 2 dimensions of motion. The fast scan direction was perpendicular to the graphene channel. The top gate voltage was held at zero volts for the data presented in the main text.

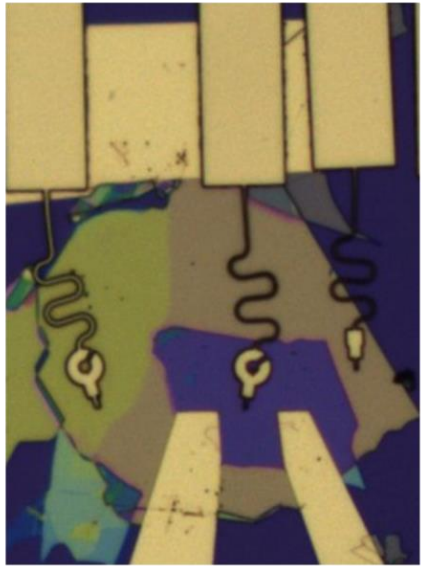

**Figure S8: hBN-encapsulated graphene device with transferred scanning top gate serpentine electrode, donut inner diameter is 3 µm.**

**Movie S1. Gold sliding on hBN at 7.6 K**
Motion amplitude is approximately 10 µm vertically, 10 µm horizontally, and 14 µm diagonally.

**Movie S2. Gold accordions longitudinal sliding at 3 Hz on hBN at room temperature**
Oscillation amplitude is approximately 2 µm.

**Movie S3. Gold accordions transverse sliding at 3 Hz on hBN at room temperature**
Oscillation amplitude is approximately 2 µm.

**Movie S4. Encapsulated graphene hockey puck *in-situ* sliding and measurement**
Measurements performed at room temperature with an amplitude of motion is 1.2 µm.
Resistance data is collected every 10 nm.